\documentclass{ws-procs9x6}

\begin{document}

\title{Cross sections at NNLO}

\author{V. DEL~DUCA\footnote{\uppercase{R}apporteur at 
\uppercase{D}\uppercase{I}\uppercase{S}06}}

\address{INFN, sez. di Torino,\\  
via P. Giuria, 1 - 10125 Torino, Italy\\
E-mail: delduca@to.infn.it}

\author{G. SOMOGYI AND Z. TR\'OCS\'ANYI}

\address{Institute of Nuclear Research of  
the Hungarian Academy of Sciences\\ and  
University of Debrecen,\\ H-4001 Debrecen, PO Box 51, Hungary\\
E-mail: z.trocsanyi@atomki.hu and somogyga@dragon.unideb.hu}

\maketitle

\abstracts{
In this talk we report on the state of the art on the calculation
of cross section at next-to-next-to-leading (NNLO) accuracy.}

\section{Higher-order calculations}

Next year, the LHC will start operating, ushering particle physics
into a completely unchartered energy realm. The LHC is a proton-proton
collider, thus in its use as a research tool it will be essential
to have the best possible theoretical understanding of QCD,
the  theory  of  the strong  interactions within the Standard Model. 
Because QCD is asymptotically free, at high $Q^2$ any cross section
can be expressed as a series expansion in $\alpha_S$. For most processes,
it suffices to evaluate the series at next-to-leading (NLO) accuracy,
which has several desirable features:
$a)$ the jet structure. At leading order it is trivial because each 
parton becomes a jet, 
at NLO the final-state collinear radiation allows up to two partons to
enter a jet; $b)$ a more refined p.d.f. evolution through the initial-state
collinear radiation; $c)$ the opening of new channels, through the inclusion
of parton sub-processes which are not allowed at leading order; $d)$ a
reduced sensitivity to the renormalisation and factorisation scales,
which are fictitious input scales,
allows to predict the normalisation of physical observables, which
is usually not accurate at leading order. That is the first step toward
precision measurements in general, and in particular toward an accurate
estimate of signal and background for Higgs and New Physics at the LHC;
$e)$ finally, the matching with a parton-shower MC generator, like
MC@NLO, which allows for a reliable normalisation of the event,
while generating a realistic event set up through showering and hadronisation.

\section{The NNLO world}

The NLO corrections, though, might be not accurate enough.  
For instance, $i)$ in the extraction of $\alpha_S$ from the data, where   
in order to avoid that the main source of uncertainty be due to the   
NLO evaluation of some production rates, like the event shapes of jet  
production in $e^+e^-$ collisions, only observables evaluated at
NNLO accuracy are considered~\cite{Bethke:2004uy};   
$ii)$ in open $b$-quark production at the Tevatron,  
where the NLO uncertainty bands are too large to test the   
theory~\cite{Cacciari:2003uh} {\it vs.} the data~\cite{Acosta:2004yw};  
$iii)$ in Higgs production from gluon fusion in hadron collisions, 
where it is known  
that the NLO corrections are large~\cite{Graudenz:1992pv,Spira:1995rr},  
while the NNLO   
corrections~\cite{Harlander:2002wh,Anastasiou:2002yz,Ravindran:2003um},   
which have been evaluated in the large-$m_t$ limit,  
display a modest increase, of the order of less than 20\%, with respect to   
the NLO evaluation; $iv)$
in Drell-Yan productions of $W$ and $Z$ vector bosons at the LHC,  
which can be used as ``standard candles'' to measure the parton luminosity  
at the LHC~\cite{Dittmar:1997md,Khoze:2000db,Giele:2001ms,Frixione:2004us}.  

In some cases, most notably in Higgs production from gluon fusion,
the central value of a prediction may change when going from NLO to NNLO.
However, the main benefit is in the
reduction of the theory uncertainty band, due to the lesser
sensitivity of the NNLO calculations to the $\mu_R, \mu_F$ scales. 
In addition, up to three partons make up the jet structure.
Thus, a lot of theoretical activity has been directed in the
last years toward the calculation of cross sections at NNLO 
accuracy~\footnote{For consistency, also the p.d.f. evolution 
has been computed to the same accuracy\cite{Moch:2004pa}}.
The total cross section~\cite{Harlander:2002wh,Hamberg:1990np}, the
rapidity distribution~\cite{Anastasiou:2003yy,Anastasiou:2003ds}
and the differential cross section~\cite{Melnikov:2006di} for Drell-Yan $W, Z$ 
production are known at NNLO accuracy. So are the total cross 
section~\cite{Harlander:2002wh,Anastasiou:2002yz,Ravindran:2003um}, 
the rapidity and the differential distributions~\cite{Anastasiou:2004xq}
for Higgs production via gluon-gluon
fusion, in the large-$m_t$ limit. However, only the calculations of
Ref.~\cite{Anastasiou:2004xq}, which has been extended to include the
di-photon background~\cite{Anastasiou:2005qj}, and Ref.~\cite{Melnikov:2006di}
allow the use of arbitrary selection cuts.

There are essentially three ways of computing the NNLO corrections: 

\noindent $a)$ Analytic integration, which is the first method to have been 
used~\cite{Hamberg:1990np}, and may include a limited
class of acceptance cuts by modelling cuts as 
``propagators''~\cite{Anastasiou:2003yy,Anastasiou:2002wq}.
Besides total cross sections, it has 
been used to produce the Drell-Yan rapidity 
distribution~\cite{Anastasiou:2003yy,Anastasiou:2003ds}.

\noindent $b)$ Sector decomposition, which is flexible enough to include any
acceptance 
cuts~\cite{Roth:1996pd,Binoth:2000ps,Heinrich:2002rc,Anastasiou:2003gr}, 
and has been used to produce the NNLO differential rates of 
Refs.~\cite{Melnikov:2006di,Anastasiou:2004xq,Anastasiou:2005qj}
and of $e^+e^-\to 2$~jets~\cite{Anastasiou:2004qd}.
The cancellation of the IR divergences is performed numerically.

\noindent $c)$ Subtraction, for which the cancellation of the divergences is  
organised in a process-independent way by exploiting the universal  
structure of the IR divergences.
However, the cancellation of the IR divergences at NNLO is very 
intricate~\cite{Kosower:2002su,Weinzierl:2003fx,Gehrmann-DeRidder:2003bm,Gehrmann-DeRidder:2004tv,Gehrmann-DeRidder:2005hi,Frixione:2004is,Somogyi:2005xz,Gehrmann-DeRidder:2005cm,Weinzierl:2006ij}, and except for test cases
like $e^+e^-\to 2$~jets~\cite{Gehrmann-DeRidder:2004tv,Weinzierl:2006ij} and 
for parts of $e^+e^-\to 3$~jets~\cite{Gehrmann-DeRidder:2005cm}, 
no NNLO numerical code has been devised yet.
The standard approach of subtraction to NNLO relies on defining approximate  
cross sections which match the singular behaviour of the QCD cross  
sections in all the relevant unresolved limits.   
For processes without coloured partons in the initial state,
in Ref.~\cite{Somogyi:2005xz}
we disentangled the various kinematical singularities of the squared matrix 
element in all singly- and doubly-unresolved parts of the phase space, 
which allows for the definition of subtraction terms for processes with any 
number of final-state coloured partons.

\section*{Acknowledgments}
VDD thanks the organizers of DIS06 for their kind hospitality and support.

\end{document}